\documentclass[12pt,a4paper,final]{iopart}

\usepackage{iopams}  
\usepackage{graphicx}
\usepackage{color}
\usepackage[breaklinks=true,colorlinks=false,linkcolor=blue,urlcolor=blue,citecolor=blue]{hyperref}
\usepackage{cite}

\graphicspath{{figs/},{./}}
\DeclareGraphicsExtensions{.pdf, .eps, .png, .jpg}

\newcommand{\eu}{\mathrm{e}}
\newcommand{\iu}{\mathrm{i}}

\begin{document}

\title[Obtaining GSE spectral statistics by breaking ergodicity]{GSE spectra  in uni-directional quantum  systems}

\author{Maram Akila}
\address{Faculty of Physics, University Duisburg-Essen, Lotharstr. 1, 47048 Duisburg, Germany}
\ead{maram.akila@uni-due.de}

\author{Boris Gutkin}
\address{Department of Applied Mathematics, Holon Institute of Technology, 58102 Holon, Israel}
\ead{boris.gutkin@uni-due.de}

\begin{abstract}
 Generically,  spectral statistics of   spinless   systems with time reversal invariance (TRI)  and chaotic dynamics are well described by the Gaussian  Orthogonal ensemble (GOE). However,  if  an additional symmetry is present, the spectrum  can be split into independent  sectors which  statistics depend  on the type of the group's irreducible representation. In particular,   this allows the construction of TRI quantum graphs with spectral statistics, characteristic  of the Gaussian Symplectic  ensembles (GSE). To this end one usually has to    use  groups admitting  pseudo-real  irreducible representations. In this paper we show how GSE  spectral statistics can be realized in TRI systems with simpler   symmetry groups lacking pseudo-real representations.  As an application, we provide  a class  of  quantum graphs with only $C_4$ rotational  symmetry possessing GSE spectral statistics.
\end{abstract}

\pacs{02.10.Ox, 02.70.Hm, 03.65.Ge, 03.65.Sq, 05.45.Gg, 05.45.Mt}
\vspace{2pc}

\noindent{\it Keywords}: 
quantum chaos, spectral statistics, random matrix theory, quantum graphs

\submitto{\JPA}

\section{Introduction}
Starting from  the  ground breaking work of E. Wigner,  \cite{Wigner},  Random Matrix Theory (RMT) has been extensively used  to describe statistics of energy levels in complex quantum  systems. But, the  scope of RMT turned out to be  much wider. With the initial development of numerical simulations in the 80's  it was realized that  RMT can be   applied equally to simple single particle systems, like quantum billiards, provided  their classical dynamics are fully chaotic  \cite{BGS}. Accordingly, the energy level statistics  of such   systems  fall in one of the three universality  classes of RMT, given by the Gaussian Unitary (GUE), Gaussian Orthogonal (GOE) and Gaussian Symplectic Ensembles (GSE) of random matrices. The relevant symmetry class is determined  by the system's behavior under  the time reversal operation  $T$. In the absence of  any additional symmetries, the spectral statistics  of systems with time reversal invariance (TRI) follow GOE if $T^2=1$ and GSE if $T^2=-1$.  If, on the other hand,  TRI is broken, then  GUE statistics  are observed \cite{Dyson}.   

For spin-less particles with intact TRI one always has $T^2=1$, which generically results in GOE statistics for the corresponding  Hamiltonians.
This, however, might change  if an additional symmetry $G$ is present in the system. In such a case  the energy spectrum   can be split into uncorrelated sectors in accordance to the  irreducible representations of $G$.
Provided  the underlying classical dynamics are chaotic, the type of representation determines the spectral statistics of the corresponding sector \cite{ Robnik, Seligman, KR97, Crys0, Gu11}. As a result, GSE statistics appear for systems with symmetry groups  possessing pseudo-real representations. This allows, for instance, the  construction of  quantum graphs with GSE statistics utilizing Cayley graphs of the quaternion group $Q_8$, \cite{Crys1, Crys2}. However,  finite isometry groups of the Euclidean space  (i.e.,  point groups)  have no pseudo-real representation. This leaves open the question whether planar  quantum billiards or  even planar quantum graphs with GSE spectral statistics can be constructed. 

\begin{figure}
 \hfill
 a)\includegraphics[height=4.1cm]{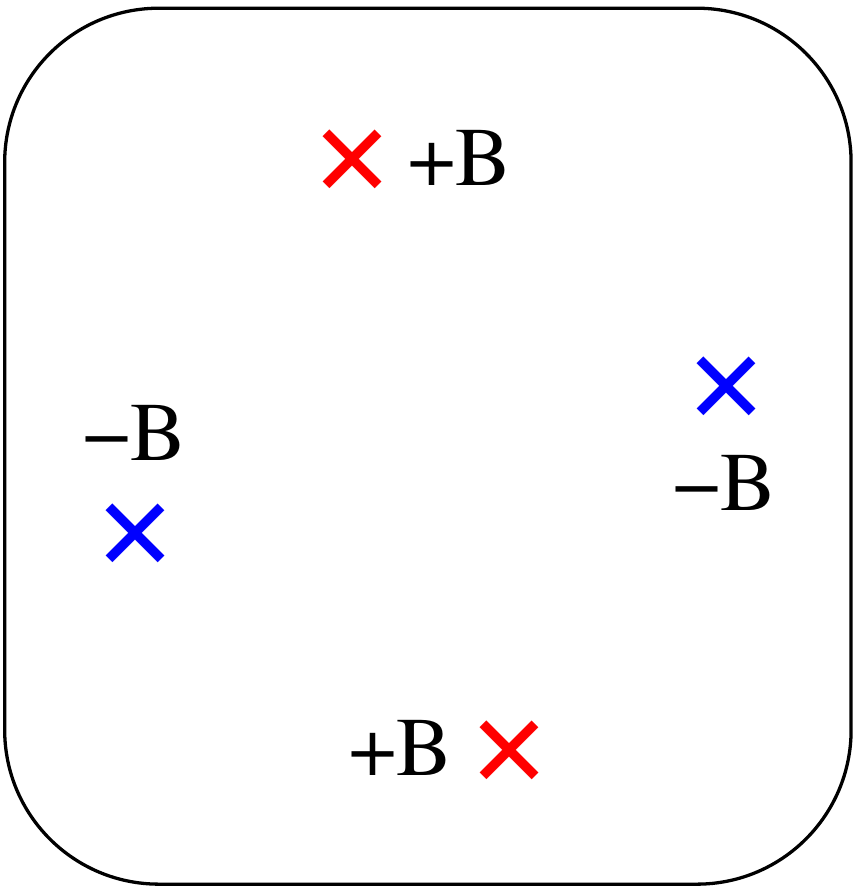} \hfill
 b)\includegraphics[height=4.1cm]{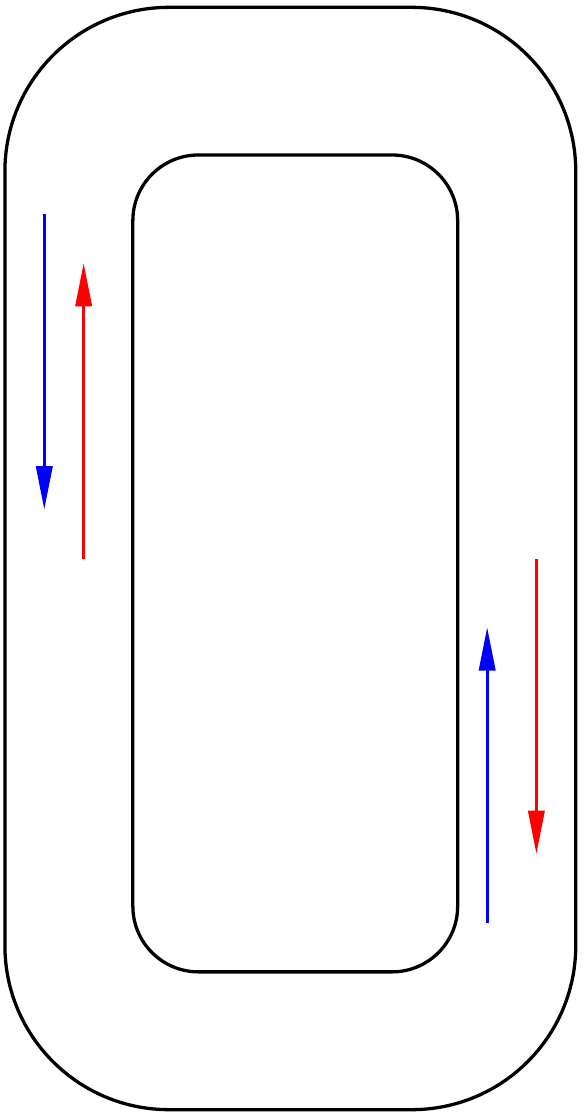} \hfill
 c) \includegraphics[height=4.1cm]{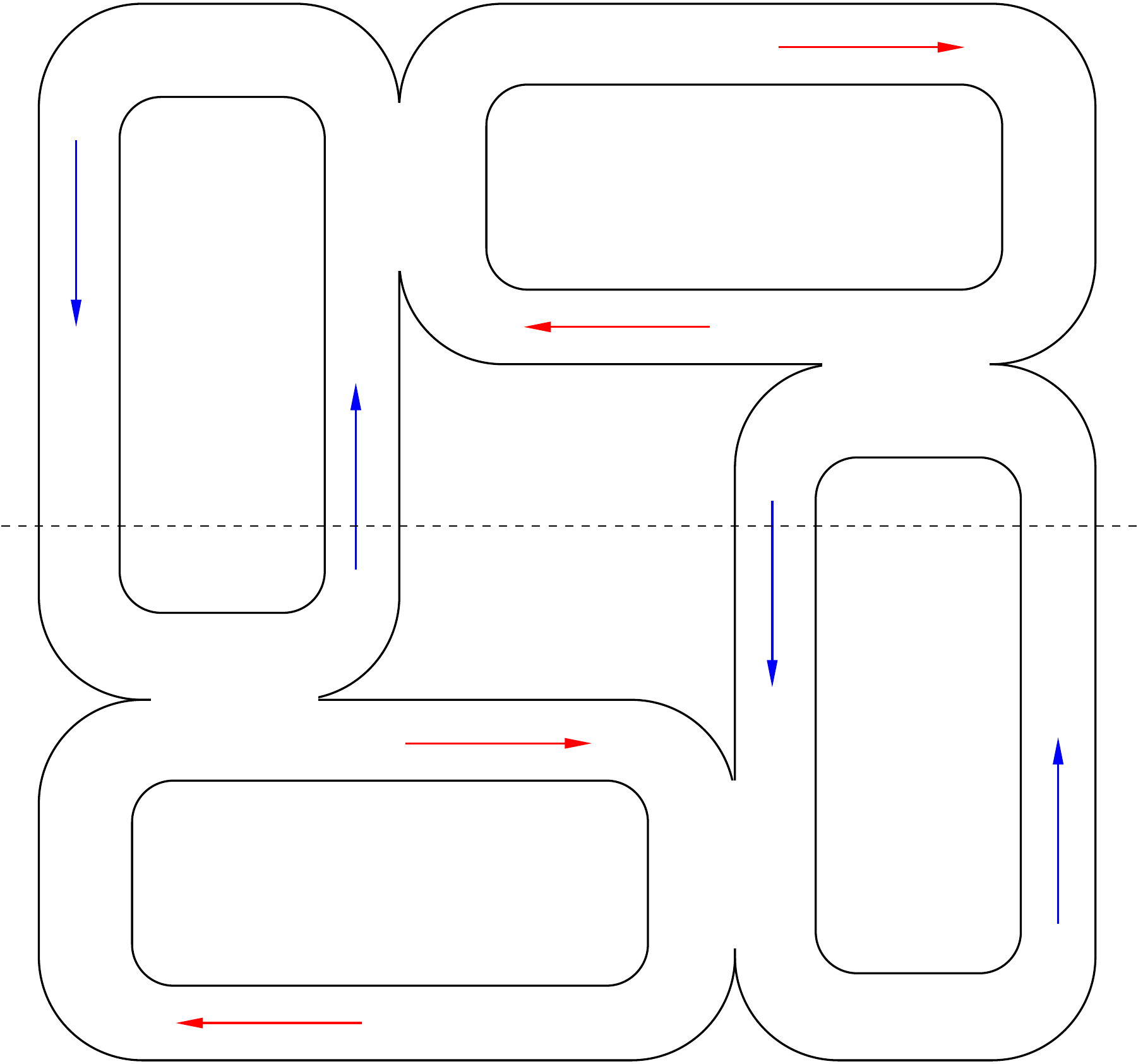}
 \caption{ a) GSE   billiard  with $\pi/2$ rotational   symmetry from \cite{Winter}. At four corners a magnetic field of opposite sign is added. b) A unidirectional billiard of constant width with chaotic dynamics. The two directions of motion illustrated  by red and blue arrows are dynamically decoupled from each other \cite{Me2}.  c) A unidirectional billiard with $C_4$ symmetry composed of four blocks in figure (b).  }
 \label{fig:sFTT1}
 \label{fig:billiardsSchema}
\end{figure}

An alternative way to obtain GSE statistics in  spin-less systems  has been recently suggested  in \cite{Winter, SieberWinter}.  The idea is that  systems with broken TRI  might posses GSE statistics  even for  simple symmetry groups,  such as $C_{2n}$,  provided that half of the group elements admit an anti-unitary representation. An example of a magnetic billiard with this property  is shown in fig.~\ref{fig:sFTT1}a. Here, the system's symmetry group  is generated by a $\pi/2$-rotation combined with the time reversal operation $T$. The $C_2$ subgroup  of rotation by $\pi$ naturally admits a unitary representation such that the whole billiard spectrum can be split into two sectors.  It can then be shown on the basis of co-representation theory \cite{WignerBook,shaw,bradley} that the two sectors posses GOE and GSE spectral statistics, respectively \cite{Winter,SieberWinter}.

The aim of this paper is to demonstrate that the above  construction can   be  carried over  to  certain types of systems with TRI, as well. Specifically, we provide an example of an unidirectional TRI  billiard and  graph  with   $C_4$ symmetry exhibiting GSE spectral statistics in the remainder. 

\section{ GSE billiards}

\subsection{GSE billiard with broken TRI}
Before considering TRI systems, let us first  explain  the construction of \cite{Winter} in some detail. The symmetry group $G$ of the billiard  in  fig.~\ref{fig:sFTT1}a  can be decomposed into the union
\begin{equation}
      G=G_+ \cup \bar{T} G_+\,,\label{thegroup}
\end{equation}
where $G_+\cong  C_2$ is the rotation group by $\pi$ and $\bar{T}=T g_{\pi/2}$ is the composition of a rotation  by $\pi/2$ with the time reversal operation $T$.  For  groups   admitting such a decomposition    ($\mathbb{Z}_2$ graded groups) Wigner developed a co-representation theory. According to it, a co-representation $\rho(g)$ of a group element $g$ is a  linear operator  if $g\in G_+$ and an antilinear  operator  if $g\in \bar T G_+$.  As in conventional representation theory, there are three types of irreducible  co-representations, which can be distinguished by  Bargmann's indicator:    
\begin{equation}
     \sigma_{\mbox{\scriptsize B}}(\rho)=\frac{2}{|G|}\sum_{g\in T G_+}\chi_\rho(g^2)\,,
\end{equation}
where  $\chi_\rho$ is the character of $\rho$ and $g^2\in G_+$ \cite{bradley}. The co-representation is real if $\sigma_{\mbox{\scriptsize B}}(\rho)=1$, complex if $\sigma_{\mbox{\scriptsize B}}(\rho)=0$ and quaternionic if $\sigma_{\mbox{\scriptsize B}}(\rho)=-1$ \cite{He37}.
Given a Hamiltonian with a  $\mathbb{Z}_2$-graded  symmetry group $G$ its spectrum can be decomposed into subspectra  corresponding  to the irreducible co-representations of $G$. Furthermore, for systems with chaotic dynamics Dyson's threefold way principle states that the spectral statistics depend on  the type of the co-representation. In particular, GSE spectral statistics occurs if $G$ admits a quaternionic  co-representation \cite{Dyson}. A direct calculation  of Bargmann's indicator for the group (\ref{thegroup}) shows that
it has  two irreducible co-representations with $\sigma_{\mbox{\scriptsize B}}(\rho)=+1$ and $\sigma_{\mbox{\scriptsize B}}(\rho)=-1$,  respectively \cite{Winter}.
Thereby, the billiard in fig.~\ref{fig:sFTT1}a has GOE and GSE subspectra corresponding to the above co-representations.

An alternative way to see the same result uses the $C_2$ subgroup of $G$ to split the billiard spectrum into a symmetric and an antisymmetric part. The role of the time reversal operation is then fulfilled by the antiunitary operator  $\bar{T}$. It is easy  to see that its  square, $\bar{T}^2$,  acts as $1$ within the symmetric sector and as $-1$  in the antisymmetric one. Accordingly, the two sectors possess GOE and GSE spectral statistics, respectively.

\subsection{GSE billiard with  TRI}
\label{Sec:GSEbilliard}
A prototype construction of a GSE billiard with TRI is sketched in fig.~\ref{fig:sFTT1}c. The billiard is composed of four blocks (shown in  fig.~\ref{fig:sFTT1}b)  such that the  total system has $C_4$ symmetry.  Being of constant width each block possesses unidirectional dynamics, i.e., clockwise and anticlockwise motions are ergodically separated \cite{BG1}. Despite of this,  the dynamics  are  fully hyperbolic such that the  Lyapunov exponent is positive almost everywhere \cite{Me2}. The resulting total billiard  possesses  unidirectional dynamics as well, but the direction of motion is interchanged between neighboring blocks. Due to the last property a rotation by $\pi/2$ of the whole billiard  changes the direction of  motion to the opposite one, i.e, switches between the two ergodic components.

To see the implications of the unidirectional dynamics  on the quantum   level it is instructive to split  the Hilbert space of the system  into clockwise and anticlockwise  sectors. After defining a chiral basis within each sector the total Hamiltonian  of the time reversal system can be cast into  the form  
\begin{equation}
    H=\left( \begin{array}{ccc}
H_+ & W^\dagger  \\
W & H_-  \end{array} \right), \qquad H_+=H_-^*,\qquad W= W^*, \label{billHam}
\end{equation} 
where  $H_+$ and  its complex conjugate $H_-$  act within  the clockwise  and  anti-clockwise sectors $\mathcal{H}_\pm$ of the Hilbert space, respectively, while $W$, $W^\dagger$  couple  between them. For strictly unidirectional systems (such as circle billiards) $W=0$ holds and their spectrum is decomposed  into  doubly degenerate eigenvalues provided by the spectra of $H_+$ and $H_-$.\footnote{Strictly speaking, for billiards of constant  width  a more  accurate model of the system  Hamiltonian (\ref{billHam})    should include, in addition, a  neutral part  $H_0=H_0^\dagger$  corresponding  to the sector $\mathcal{H}_0$ of the Hilbert space spanned by bouncing ball modes \cite{Darm}.  For instance, in the circle billiards   Bessel functions provide a sequence of eigenstates with zero angular momentum, i.e., they belong to $\mathcal{H}_0$. However, the ratio between the dimensions of $H_0$ and $H_\pm$ tends to $0$  in the semiclassical limit. For the sake of simplicity of exposition we omit the neutral sector from the Hamiltonian  (\ref{billHam}).} In the absence of any additional symmetry $H_\pm$ has the form of a generic Hermitian matrix. Therefore the doublets   of chaotic   unidirectional  systems  possess GUE spectral statistics.  But for unidirectional systems with $C_4$ rotation symmetry as in fig.~\ref{fig:sFTT1}c,  $H_\pm$ possesses an additional constrain. Namely,  under rotation by $\pi/2$,   $H_+$ is transformed   into  $H_-$. This makes the combination of $ g_{\pi/2}$  with the time   reversal operation,
$T g_{\pi/2}$, a symmetry of  $H_\pm$. In other words, $H_\pm$ stays invariant under the action of the group (\ref{thegroup}).  Accordingly, for strictly unidirectional systems with $C_4$ symmetry and chaotic dynamics one might expect a spectrum of doublets distributed according to GOE  statistics and quadruplets obeying GSE statistics. 

 For  the billiards shown in fig.~\ref{fig:sFTT1}b,c  unidirectionality is,  in fact, weakly broken on the quantum level
via diffraction effects, i.e., $W\neq 0$ is reminiscent of singular perturbations \cite{MeMaram}. In particular,  this lifts the degeneracies and thereby affects the spectral statistics of the system.  
Nevertheless,   traces   of  GUE   statistics have  been previously  observed in  billiards of the same  type as presented in  fig.~\ref{fig:sFTT1}b,  \cite{Prosen}. And in the same spirit it might be expected that traces of  GSE and GOE spectral statistics (depending on the sector) are  recognizable in the billiard depicted  in fig.~\ref{fig:sFTT1}c.  To avoid these complications we illustrate the above construction for a  more accessible   class of systems -- quantum graphs, where unidirectionality is exact.

\begin{figure}
\hfill
 a)\includegraphics[height=3.1cm]{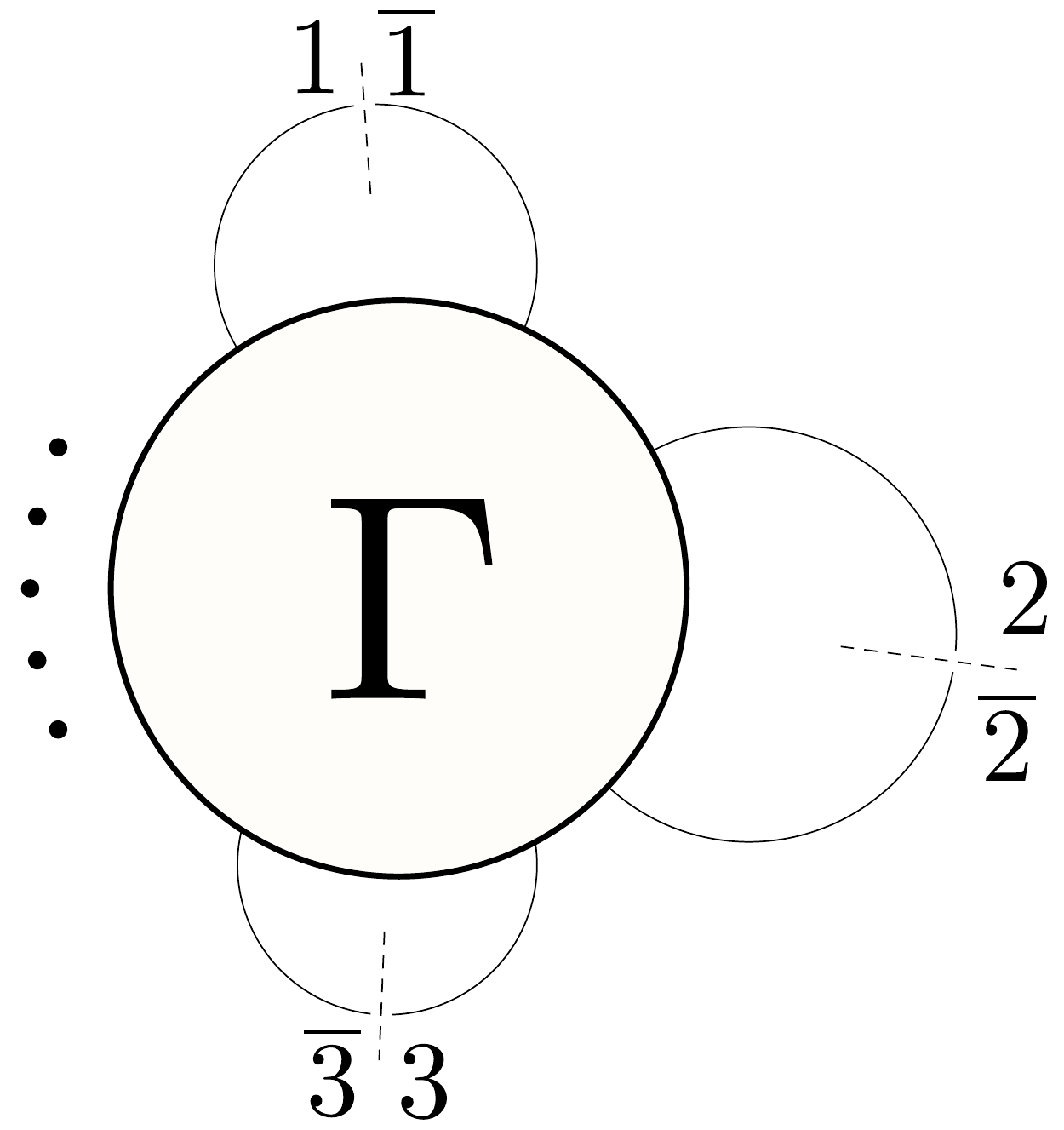} 
 \hfill
  b)\,\,\includegraphics[height=3.1cm]{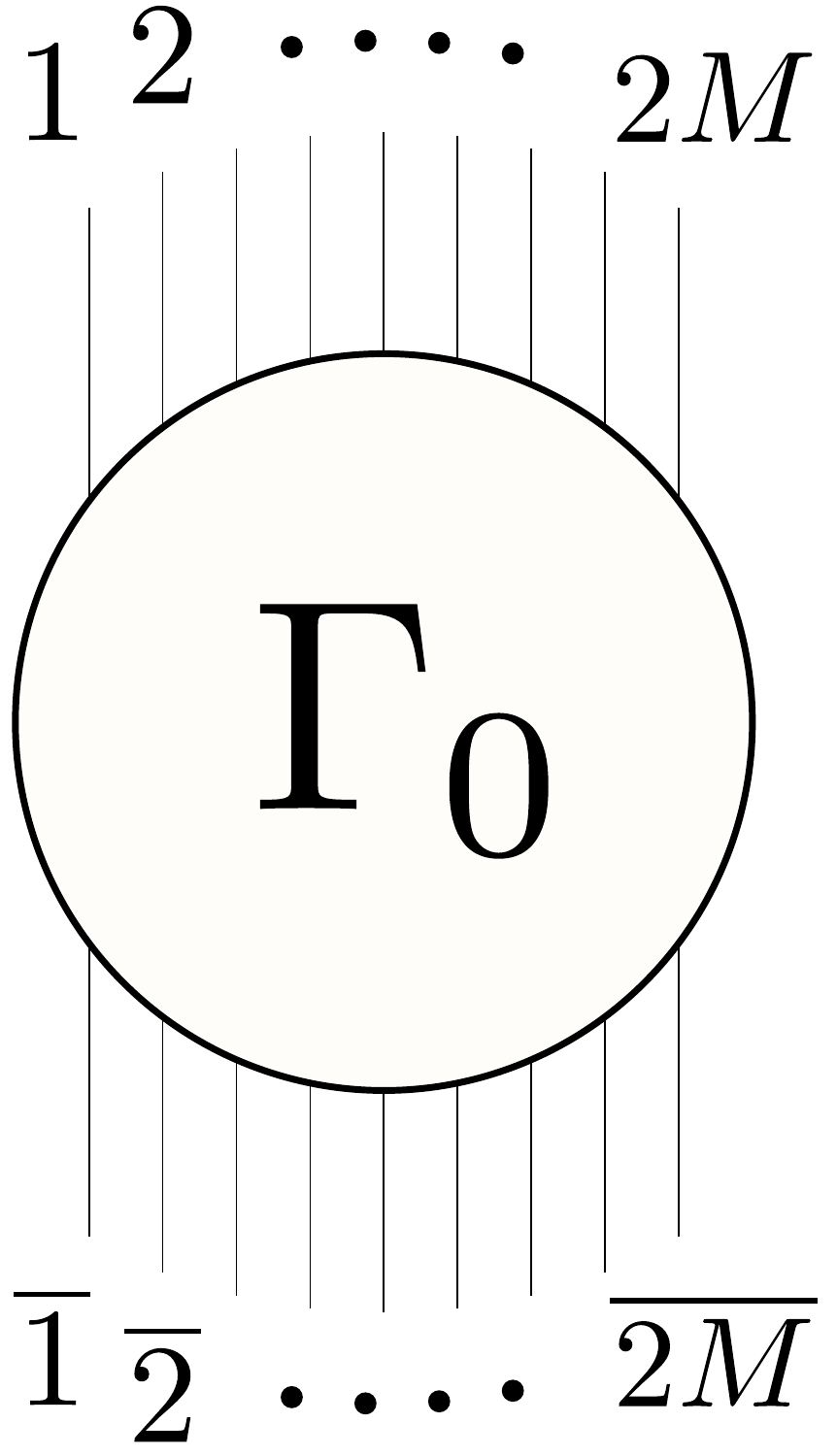}
  \hfill
  c)\includegraphics[height=3.1cm]{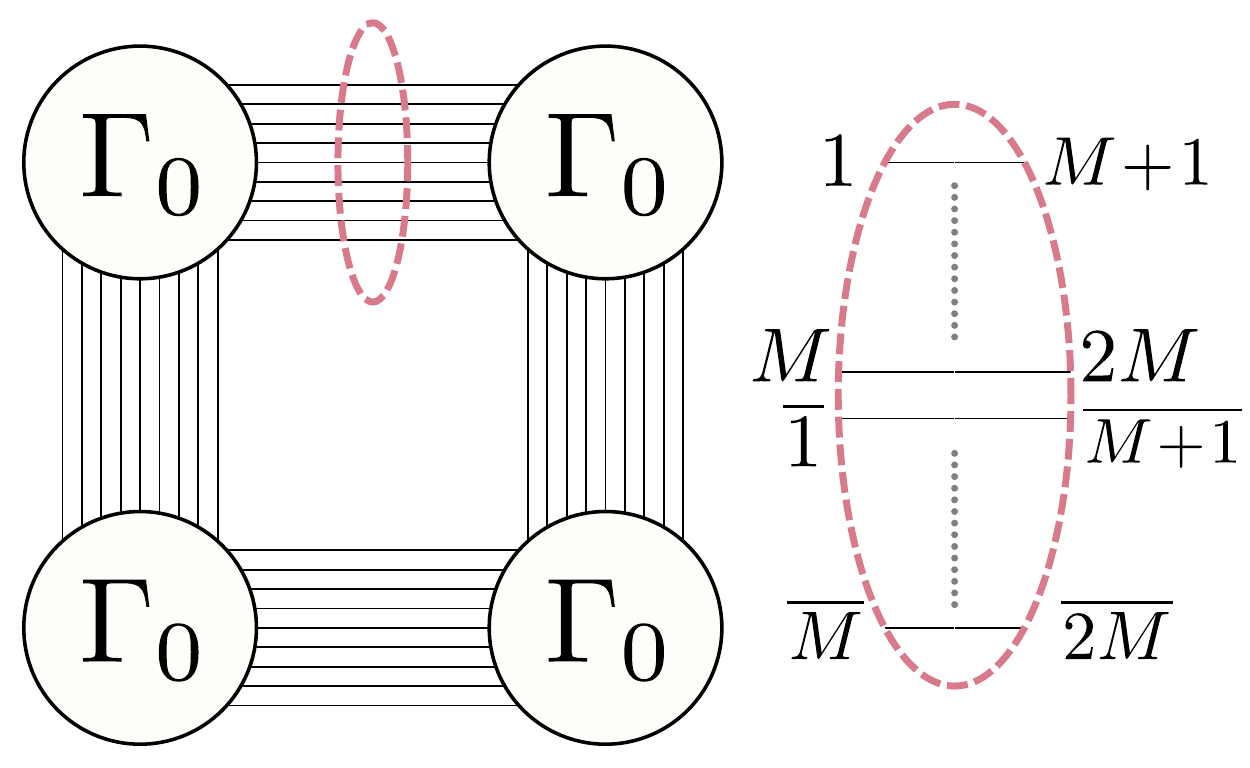}
 \caption{a) A unidirectional graph $\Gamma$.  b) The corresponding open graph $\Gamma_0$ with $2M$ entering/exiting legs. Its  scattering matrix satisfies conditions  $S^{(0)}_{i,j}=S^{(0)}_{\overline{i},\overline{j}}=0$. c) Unidirectional GSE  quantum graph which is composed of four blocks $\Gamma_0$. Different  entering/exiting legs of  ${\Gamma}_0$  are reconnected according to the rule   $i\leftrightarrow  M+i$, $\overline{i}\leftrightarrow  \overline{M+i}, \quad i\in \{1,\dots M\}$.   }
 \label{fig:sFTT2}
 \label{fig:graphsSchema}
\end{figure}

\section{Unidirectional  GSE graphs}

Based on the prototype GSE billiard shown in fig.~\ref{fig:sFTT1}b we construct a family of quantum graphs with GSE spectral statistics. The basic idea is to substitute each block  in fig.~\ref{fig:sFTT1}c with a quantum graph possessing unidirectional dynamics. 

\subsection{Unidirectional  quantum graphs.}
Let  $\Gamma$ be a graph   with   $N$ bonds  connecting  $V$ vertices.
Then the corresponding quantum graph is defined in terms of the Schr{\"o}dinger equation on the bonds, i.e., free wave propagation, satisfying  some  boundary conditions at the respective vertices  \cite{KotSmil}.
To formulate this problem in a way more suitable for our discussion we introduce the corresponding quantum map acting on directed graph bonds. To this end each bond of $\Gamma$  is equipped with  two directions  such that the set  of $2N$ directed bonds can be split into two halves, $1_+,\dots, N_+$  and    ${1}_-,\dots, {N}_-$, where $i_+$ and $ i_-$ correspond to the opposite directions  of  the $i$'th bond of the graph $\Gamma$. We  denote with $\Gamma_+$,  $\Gamma_-$  the pair of  graphs with the same topological structure as $\Gamma$, but  composed of the directed bonds  $1_+,\dots, N_+$  and    ${1}_-,\dots, {N}_+$,  
respectively.\footnote{As the assignment of $+$ or $-$ to a particular direction of the graph's bonds is a matter  of  choice,  there are $2^N$   different ways to define $\Gamma_\pm$ for a general graph. However, for unidirectional graphs the choices are more restricted, as detailed in the remainder of the text.
}
For a given wavenumber $k$ the quantum map    is defined as a product, $S(k)=S_0 \Lambda(k)$, of the diagonal part, 
\begin{equation}
	\Lambda(k)={\mbox{diag}}\{\eu^{\iu k\ell_1}, \dots, \eu^{\iu k\ell_N}, \eu^{\iu k\ell_1}, \dots, \eu^{\iu k\ell_N}\}\,,
\end{equation}
depending on the bond length's $\ell_i$ and the constant scattering matrix $S_0$ which   encodes the boundary conditions at  the vertices. Specifically,   $S_0$ is composed of  local unitary blocks $\sigma^{(v)}, \, v=1,2,\dots, V$,  describing  the  scattering processes  at the  vertex $v$.   The quantum map $S(k)$ acts on the $2N$-dimensional vectors $(\psi_1,\dots, \psi_N,\psi_{N+1}, \dots, \psi_{2N})$, where   the first $N$ indices belong to   $\Gamma_+$, and the remaining to $\Gamma_-$.    The quantum graph spectrum,  $\lambda_n=k_n^2$, is then found as the solutions $k_n, n=1, \dots \infty$, to  the secular equation \cite{KotSmil, GutSmil}
\begin{equation}
    \det (I- S(k) )=0\,.\label{grapheq}
\end{equation}

The  unidirectional quantum graphs were introduced in \cite{MeMaram}. Their  construction  is based on a  special choice of  the  local  TRI scattering matrices,
\begin{equation}
    \vec{\psi}_{\mbox{\scriptsize out}} =\sigma^{(v)} \vec{\psi}_{\mbox{\scriptsize in}}\,,
    \label{eq:locScat}
\end{equation}
which relate incoming and outgoing wave functions $\vec{\psi}_{\mbox{\scriptsize in}}$, $\vec{\psi}_{\mbox{\scriptsize out}}$ at the vertex $v$ \cite{MeMaram}. Specifically, the matrix $\sigma^{(v)}$ must have  even dimensions  $2M_v\times 2M_v$  such that the $2M_v$ indices  can be separated into two subgroups --   ``entrances" $ \{1,2,\dots  M_v\}$  and  ``exits"  $ \{\overline{1},\overline{2},\dots,  \overline{M_v}\}$.  The corresponding   matrix elements satisfy $\sigma^{(v)}_{i,j}=\sigma^{(v)}_{\overline{i}, \overline{j}}=0$, for  $i,j \in 1,\dots, M_v$. Thereby each vertex only connects ``entrances" to ``exits" ($i\to \overline{j}$) and vice verse. 
A TRI,  unidirectional graph $\Gamma$ is then constructed by pairwise   connecting exits  to   entrances from different vertices. Due to the  dynamical separation of the two ``directions of motion"  at the graph vertices the resulting scattering matrix $S(k)$ can be split  into two blocks \cite{MeMaram},
 \begin{equation}
     S(k)=  \left( \begin{array}{ccc}
S_+ & 0  \\
0 & S_-  \end{array} \right), \qquad S_+=S_-^T\,,\label{Structure}
\label{eq:scatMatStruct}
\end{equation}
where $S_+$ and $S_-$  act on the $\Gamma_+$ and $\Gamma_-$ parts of the vectors. Accordingly, the transition  matrix $|S_{i,j}|^2$,
for  the  underlying classical Markov chain,    possesses a doubly degenerate largest eigenvalue $1$ corresponding to two ergodic components  supported on $\Gamma_+$ and $\Gamma_-$, respectively.

By eq.~(\ref{Structure})  the spectrum of a TRI  unidirectional  quantum graph is doubly degenerate, the identical eigenvalues stem from $S_+$ and $S_-$, respectively.
Furthermore, in the absence of symmetries  $S_\pm$ takes on the form of a generic unitary matrix  such that  the  doublets   $\{\lambda_n\}_{n=1}^\infty$ defined by eq.~(\ref{grapheq})  possess  GUE statistics rather than GOE, as would have been expected for a TRI systems.

\subsection{Construction of GSE quantum graphs.} 
Similar to the GSE billiard presented in fig.~\ref{fig:billiardsSchema}c its quantum graph analogue is build from four copies of a TRI unidirectional quantum graph $\Gamma$. The procedure goes as follows:
In the first step,  we cut $2M$ bonds of the original graph $\Gamma$ in the middle, turning it into the open graph $\Gamma_0$,  see  figs.~\ref{fig:sFTT2}a,b.
Due to unidirectionality the $4M$ loose ends of $\Gamma_0$  are naturally separated into two groups of   $2M$     ``exits''   and  $2M$   ``entrances", as discussed in the previous  section. We will enumerate all  ``entrances"  by numbers from  $1$ to $2M$ (in an arbitrary order)   and the respective ``exits" by   numbers with bars,  such that     $i$ and $\bar{i}$ correspond to one and the same original bond of $\Gamma$. 
Due to the unidirectionality of the original graph, the   scattering matrix  corresponding to  the open graph $\Gamma_0$   satisfies  $S^{(0)}_{i,j}=S^{(0)}_{\overline{i},\overline{j}}=0$  for  $i, j\in 1,\dots  2M$ (compare with the analogous property of  the vertex matrices $\sigma^{(v)}$). In other words,  $S^{(0)}$  connects ''entrances'' with ''exits'' and vice versa. 
In  the second step we divide all entering (resp. exiting)  ends  into two   groups   $\{1,\dots, M\}\cup \{M+1,\dots, 2M\}$  (resp. $\{\overline{1},\dots, \overline{M}\}\cup \{\overline{M+1},\dots, \overline{2M}\}$) and connect four copies of $\Gamma_0$ according to the rule  (see fig.~\ref{fig:sFTT2}c): 
\[i\leftrightarrow  M+i\,,
\qquad \bar{i}\leftrightarrow  \overline{M+i}\,,
\quad i\in \{1,\dots M\}\,.\]



The constructed quantum graph is strictly unidirectional and  its scattering matrix $S(k)$ has the same split form (\ref{eq:scatMatStruct}), where  $S_\pm$  correspond to the two different  ergodic components.  On the other hand, since the same type of entries are put together, e.g., bar with bar,  the directionality between neighbouring copies of $\Gamma$ is reversed - everything entering from $\Gamma_+$ ends in $\Gamma_-$ and vice versa. As a result, the  rotation  by $\pi/2$ leads to an exchange of the clockwise and anti-clockwise blocks, such that
\begin{equation}
    g_{\pi/2}\,S_\pm\, g^{-1}_{\pi/2}=S_\mp=S^T_\pm\,. \label{SymEq}
\end{equation}
From eq.~(\ref{SymEq})   and the  time reversal invariance  condition, $TST=S^T$,
immediately follows that $[S_\pm,Tg_{\pi/2} ]=0$, i.e., $Tg_{\pi/2}$ is the anti-unitary symmetry of $S_\pm$. 
Repeating the arguments of section \ref{Sec:GSEbilliard}, the spectra of the $S_\pm$ blocks can be split into periodic and anti-periodic sectors with respect to the rotation by $\pi$.
The resulting  spectral statistics are expected to be in good agreement with the ones of doubly degenerate GOE and four-fold degenerate GSE, respectively, provided that the underlying graph $\Gamma$ obeys (GUE) RMT predictions and the connectivity between the four copies is of sufficient rank. The topological structure of  $\Gamma$ can otherwise be arbitrary. In particular, it might also be planar. 

\begin{figure}
\hfill
 a)\includegraphics[height=5.1cm]{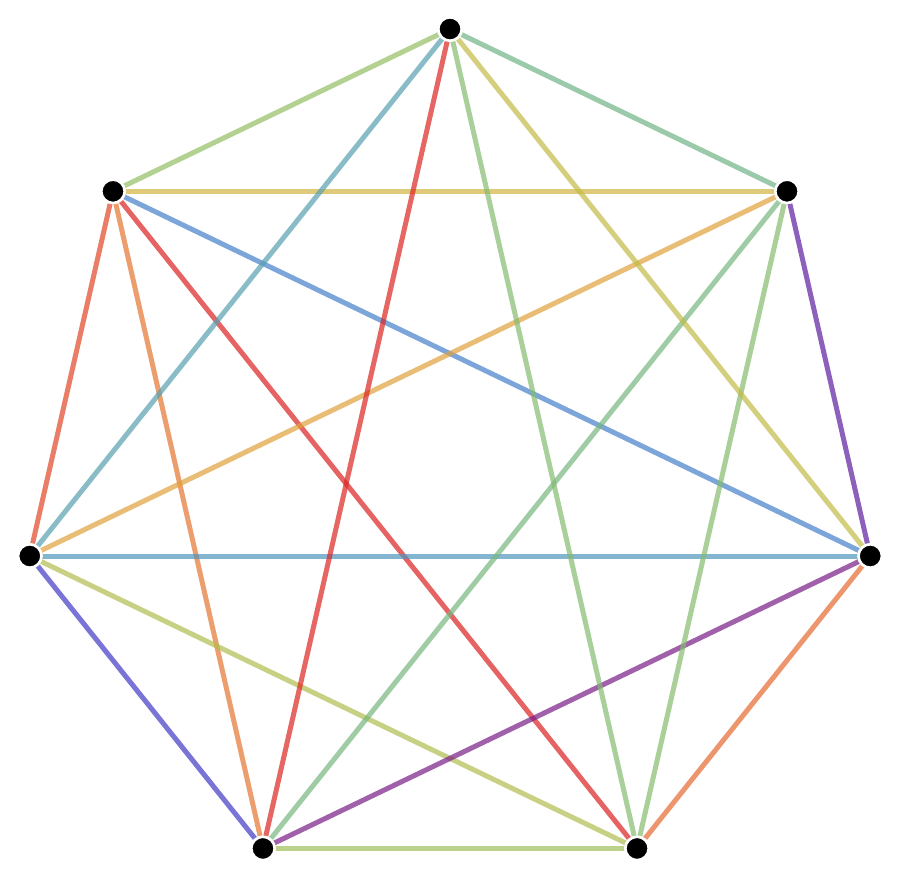}
 \hfill
  b)\includegraphics[height=5.1cm]{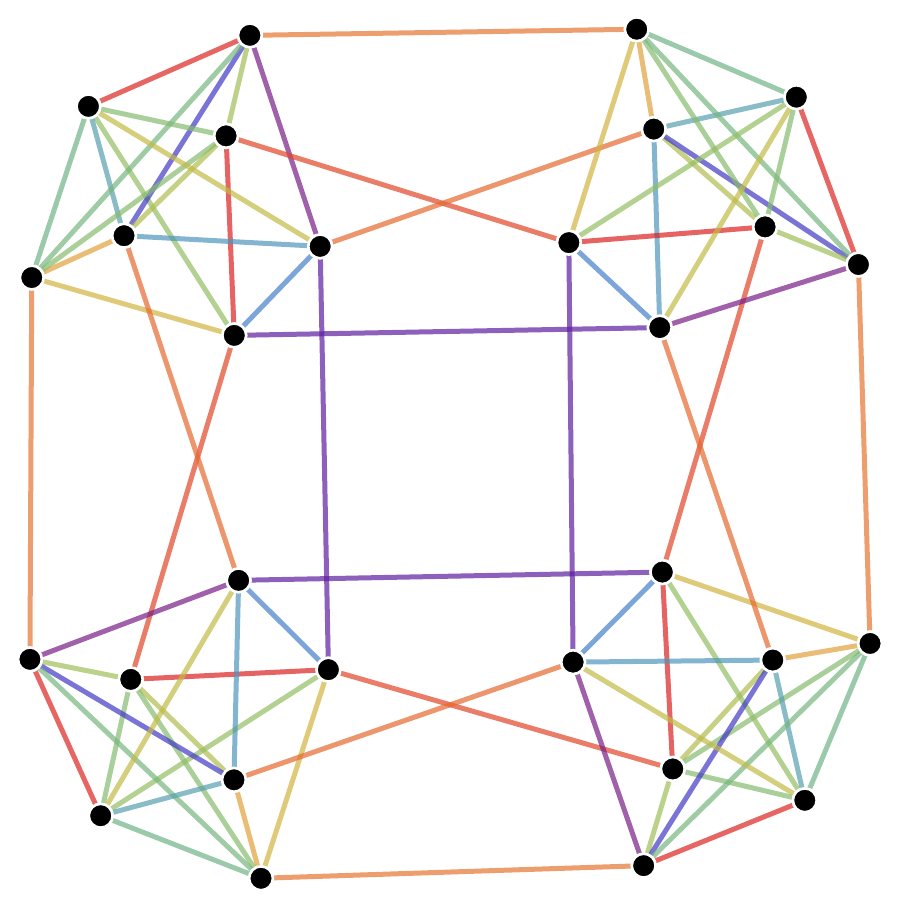}
  \hfill
 \caption{a) Fully connected unidirectional graph. b) The corresponding GSE graph constructed as in fig.~\ref{fig:graphsSchema}. Edges with similar lengths are colored similarly.}
 \label{fig:sFTT3}
\end{figure}

\begin{figure}
    \hfill
    \includegraphics[width=0.4\textwidth]{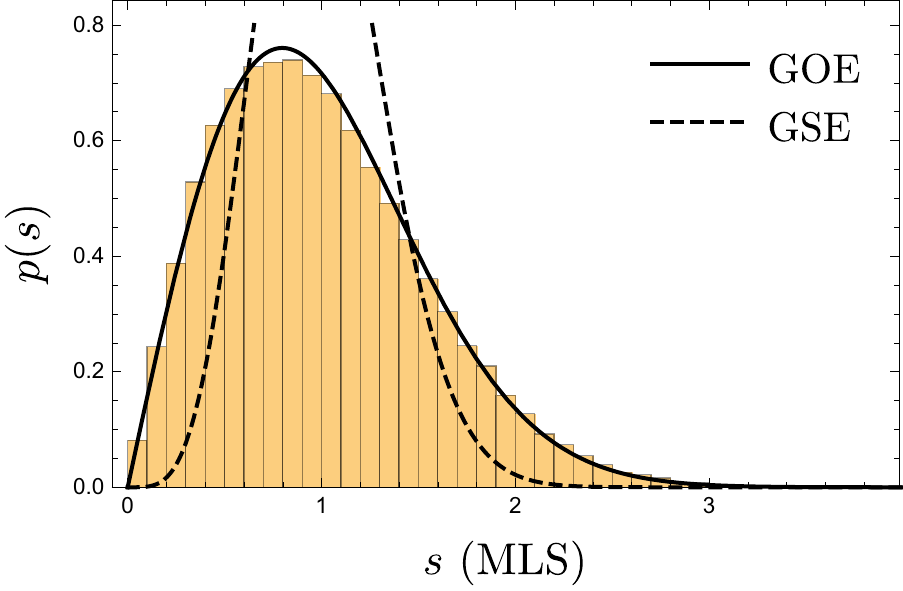}
    \hfill
    \includegraphics[width=0.4\textwidth]{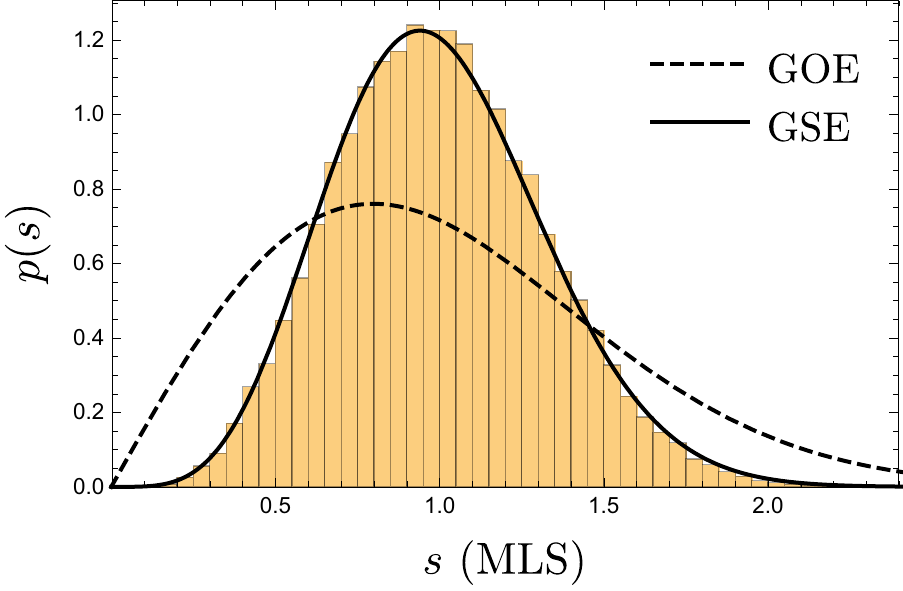}
    \hfill
    \caption[$p(s)$ for $k$-eigenvalues of CSE graph.]{Nearest neighbor spacing distributions  $p(s)$ for the first $10^5$ $k$-eigenvalues (adjusted to their respective mean level spacing) of the graph shown in figure \ref{fig:sFTT3}b. The two distributions are  separated according to  the   eigenfunctions behavior  under the graph's rotation by $\pi$. Left (resp. right)   side shows the nearest neighbour distribution for the periodic  (resp.  anti-periodic) eigenfunctions.  The lines represent the Wigner surmise for GOE and GSE.}
    \label{fig:nnSpacingK}
\end{figure}

In  fig.~\ref{fig:sFTT3}b we provide an example of a GSE quantum graph, where each block $\Gamma$ is a fully connected unidirectional graph,  shown in  fig.~\ref{fig:sFTT3}a. Four edges  of the graph $\Gamma$ are cut in the middle and reconnected  according to the above rules.  The  $2N\times 2N$ evolution matrix (quantum map) $S(k)$ is defined then, as in (\ref{grapheq}), with   $N=84$. The nearest-neighbour spacing distribution of the corresponding $k_n$ spectrum, for  the separate sectors, is presented in fig.~\ref{fig:nnSpacingK}. As predicted, it    shows an  excellent  agreement with  GOE and  GSE statistics, respectively. 




\section{Conclusions}

We provided an explicit example of TRI quantum graphs  with GSE spectral statistics. As opposed to  previous cases these systems  have a simple $C_4$ symmetry group, which does not admit a pseudo-real representation.
This  allows (at least in principle) to realize them as planar structures.  The key ingredient to our construction is the dynamical separation of the system into two ergodic components.
In combination with  the four-fold rotation symmetry this leads to two uncorrelated subspectra obeying the GOE and GSE statistics, respectively.

The construction can be  straightforwardly extended to a general $\mathbb{Z}_2$-graded group, provided it has a pseudo-real co-representation.   In particular,  the $C_4$ symmetry group can be traded off to any rotation group $C_{4n}$. The corresponding graph is constructed  by cyclically  connecting $4n$ copies of a unidirectional graph. In such a case the spectrum of the resulting  quantum graph is divided into $n$ doubly degenerate GOE subspectra and $n$ four-fold degenerate GSE subspectra. 
It would be  of interest to explore whether an experimental  microwave realization of  such  quantum graphs is possible.


\section*{Acknowledgements}
We are grateful to M. Sieber and  A. Winter  for  useful  discussions.

\end{document}